\documentstyle[preprint,aps]{revtex}
\tighten
\draft
\widetext
\input epsf
\preprint{YITP-99-52}
\bigskip
\bigskip
\begin{document}
\title{Non-commutative Unification in Brane World}
\medskip

\author{Zurab Kakushadze\footnote{E-mail: 
zurab@insti.physics.sunysb.edu}}

\bigskip
\address{C.N. Yang Institute for Theoretical Physics\\ 
State University of New York, Stony Brook, NY 11794}

\date{October 9, 1999}
\bigskip
\medskip
\maketitle

\begin{abstract}
{}We point out that in (open) string compactifications with non-zero NS-NS
$B$-field we can have large Kaluza-Klein thresholds even in the small
volume limit. In this limit the corresponding gauge theory description is in
terms of a compactification on a non-commutative space ({\em e.g.}, a torus 
or an orbifold thereof). Based on this observation we discuss a
brane world
scenario of non-commutative unification via Kaluza-Klein thresholds. In
this scenario, the unification scale can be lowered down to the TeV-range,
yet the corresponding compactification radii are smaller than the
string length. We discuss a potential application of this scenario in the
context of obtaining mixing between different chiral generations which is
{\em not} exponentially suppressed - as we point out, such mixing is
expected to be exponentially suppressed in certain setups with large volume
compactifications. We also point out that T-duality is broken by certain 
non-perturbative twisted open
string sectors which are supposed to give rise to chiral generations, so
that in the case of a small volume compactification with a rational
$B$-field we cannot T-dualize to a large volume description. In this sense, the
corresponding field theoretic picture of unification via Kaluza-Klein
thresholds in this setup is best described in the non-commutative language.

\end{abstract}

{}With the advent of D-branes \cite{polchi} it has become evident that there
might exist a logical possibility where the Standard Model gauge and matter 
fields reside inside of $p\leq 9$ spatial dimensional $p$-branes (or a set
of overlapping branes), while gravity lives in a larger (10 or 11)
dimensional bulk of space-time. This brane world picture\footnote{For
recent developments, see, {\em e.g.}, 
\cite{witt,lyk,TeV,dienes,3gen,anto,ST,3gen1,TeVphen,BW,zura}. 
The brane world picture in the effective field theory context 
was discussed in \cite{early,shif}. Large radius
compactifications were originally considered in \cite{ANT} in the context of
supersymmetry breaking.} {\em a
priori} appears to be a viable scenario \cite{BW}. One of the implications
of the brane world scenario is that the string scale $M_s$ 
can be much lower than
the four dimensional Planck scale $M_P$ 
\cite{witt,lyk,TeV}. In fact, in \cite{TeV} it was proposed that $M_s$ as
well as the fundamental (10 or 11 dimensional) Planck scale $M_{Pf}$
can be in the 
TeV range\footnote{By the TeV-range we do {\em not} necessarily mean $M_s
\sim 1~{\mbox{TeV}}$. In fact, as was argued in \cite{zura}, the gauge
coupling unification constraints seem to imply that $M_s$ cannot really be
lower than $10-100~{\mbox{TeV}}$.}. The observed weakness of the four
dimensional gravitational coupling (that is, the fact that the four
dimensional Planck scale $M_P\gg M_s,M_{Pf}$) then requires the presence of at
least two large ($\gg 1/M_s$) compact 
directions\footnote{Recently it was proposed in \cite{gogbe} that
gravity could also be ``localized'' and such compact directions might
actually not be required. For subsequent works along these lines, see, {\em
e.g.}, \cite{RS}.}
(which are transverse to
the branes on which the Standard Model fields are localized). 
A general discussion of possible 
brane world embeddings
of such a scenario was given in \cite{anto,ST,BW}. 
In \cite{TeVphen} various non-trivial phenomenological 
issues were discussed in the context of the TeV-scale 
brane world scenario, and it was argued that this possibility 
does not appear to be automatically ruled out.

{}The gauge coupling unification in such a scenario would have to arise in
a way drastically different from the usual MSSM unification. In fact, a
higher dimensional mechanism for gauge coupling unification in the
TeV-scale brane world context was proposed in \cite{dienes}. In this
mechanism the unification scale is lowered compared with the usual MSSM
unification scale if one assumes that the Standard Model fields are
localized on $p$-branes with $p>3$ (actually, $p=4$ or 5 \cite{zura}), 
and the sizes of $p-3$ compact
directions inside of these $p$-branes are somewhat large compared with
$1/M_s$. Then the evolution of the gauge couplings above the corresponding
Kaluza-Klein (KK) thresholds is no longer logarithmic but power-like \cite{TV}.
In \cite{zura} a TeV-scale Supersymmetric Standard Model (TSSM) was
proposed, where the gauge coupling unification indeed occurs via such a
higher dimensional mechanism. Moreover, the unification in the TSSM, which
occurs in the TeV-range, is as precise at one loop as in the MSSM, and it
would also explain why the unification in the MSSM is not an
accident. Furthermore, higher loop effects in the TSSM 
were shown in \cite{zura} to be subleading due to the underlying ${\cal
N}=2$ supersymmetry at the heavy KK levels. 

{}Concrete realizations of the above higher dimensional mechanism for
gauge coupling unification in the TeV-scale brane world are expected to be
non-trivial. In particular, it is unclear at present how to explicitly 
embed models such as the TSSM in the string theory framework. This, in
turn, is not so surprising as a completely 
successful embedding of, say, the MSSM in
string theory is still lacking. In part this might be due to the lack of
necessary technology for the corresponding model building as such string 
vacua are expected to be rather non-trivial. Semi-realistic ${\cal N}=1$ 
supersymmetric examples which
possess some (but not all) desirable phenomenological features (such as
three chiral generations and Standard Model-like gauge groups) were
constructed in \cite{3gen,3gen1} via Type I/orientifold
compactifications\footnote{For other
recent developments in four dimensional Type
I/orientifold compactifications, see, {\em e.g.}, \cite{TypeI,NP}.}.
However, as was pointed out in \cite{zura}, examples of this type do not
quite seem to fit in the context of unification via Kaluza-Klein
thresholds. In particular, some matter fields (such as some of the chiral
generations) must {\em not} propagate in the aforementioned extra $p-3$
dimensions or else the corresponding one-loop 
$\beta$-function coefficients for the
power-like running become too positive, and the gauge couplings blow up
before they unify. That is, in order to have a successful unification in
the TeV-range, we must assume that some (more precisely, at least one) of
the chiral generations is localized at a point in the extra $p-3$
dimensions (see the first reference in \cite{zura} for details). It is then
not difficult to see that such a model cannot be constructed in the
{\em perturbative} orientifold framework. Indeed, within this framework the
only way to have matter localized in a smaller dimensional subspace is if
it comes from open string sectors where one of the two ends of the open
string is attached to one set of D-branes, while the other end is
attached to a different set of D-branes, with these two sets of
D-branes overlapping in the subspace of the desired localization. For
instance, consider two sets of D5-branes, call them D$5_1$- and
D$5_2$-branes, in\footnote{Here we choose a simple toroidal 
compactification only for illustrative
purposes. In more realistic cases one could equally successfully consider,
say, toroidal orbifolds.}
 ${\bf R}^{3,1}\otimes T^2\otimes T^2\otimes T^2$. 
Let D$5_1$-branes wrap the first $T^2$ while D$5_2$-branes wrap the second
$T^2$. Then the $5_1 5_2$ matter is localized at a point in the compact
space where these two sets of branes intersect. Now, let us assume that the
Standard Model gauge group comes from the same set of branes, say,
D$5_1$-branes. Note that this is essentially unavoidable or else we would
have no unification prediction to begin with - indeed, if different gauge
subgroups of the Standard Model gauge group come from different sets of
branes (with generically different corresponding compactification volumes),
then the gauge couplings for these subgroups at the string scale (which we
would eventually like to identify with the unification scale) are {\em not} 
supposed to be the same. So from now on we will {\em assume} that all three
subgroups of the Standard Model gauge group $SU(3)_c\otimes SU(2)_w\otimes
U(1)_Y$ come from the same set of branes, in particular, D$5_1$-branes.
If this is the case, then at least 
some of the $5_1 5_2$ matter corresponding to the Standard Model
matter fields would have to be
charged under three different gauge factors - $SU(3)_c\otimes SU(2)_w$
(here we are being overly generous with $U(1)_Y$ whose more careful
treatment could only worsen the situation) as well as a gauge subgroup of
the D$5_2$-brane gauge group. In particular, left-handed electroweak
doublet quarks would have to be such states. This is, however, impossible
in the perturbative orientifold framework as there open strings have only
two ends, hence only one type of distinct Chan-Paton factors is allowed for
a given set of D-branes.  

{}Thus, we must find other possible mechanisms for matter localization which
would take us outside of the perturbative orientifold construction. In
\cite{zura} it was pointed out that an adequate framework for such model
building might be that of {\em non-perturbative} orientifolds examples of
which have been recently constructed in \cite{NP}. In such vacua, which 
(generically) are
non-perturbative from both Type I and heterotic viewpoints, there are
additional {\em twisted} open string sectors (which do not possess
perturbative orientifold description) giving rise to states charged under,
say, the gauge group coming from some set of D5-branes and, at the same
time, localized at orbifold fixed points. This is analogous to what happens
in perturbative heterotic superstring where some of the states charged
under the appropriate subgroups of ${\mbox{Spin}}(32)/{\bf Z}_2$ or
$E_8\otimes E_8$ come from twisted closed string sectors. In fact, this is
not just a mere analogy but has more substance as using Type I-heterotic
duality one can in certain cases map twisted open string sectors to twisted
closed string sectors which have perturbative heterotic description
\cite{NP}. In more complicated cases where no perturbative description
exists in either pictures one can still understand such states using a
``hybrid'' approach which involves utilizing some geometric aspects of
orbifold compactifications.   

{}As we have already mentioned, the matter localized at orbifold fixed
points in non-perturbative orientifold compactifications essentially
behaves as twisted closed string states in heterotic string theory. In
particular, couplings of such twisted open string matter fields localized
at orbifold fixed points obey the point-group and space-group selection
rules (see, {\em e.g.}, \cite{space} for details). In the following we will
mostly be interested in the space-group selection rules, so let us briefly 
review their essence using a simple example.  

{}Thus, consider a $T^2/{\bf Z}_3$ orbifold, where the generator $g$ of the
orbifold group acts crystallographically on $T^2$ by $2\pi/3$
rotations. The metric on $T^2$ is given by $g_{ab}=(v/\sqrt{3}) 
e_a\cdot e_b$,
$a,b=1,2$, where $v$ is the volume of $T^2$, and  
$e_a$ are the vectors generating the $SU(3)$ root lattice
$\Lambda\equiv\{n^a e_a| n^a\in {\bf Z}\}$ (note that $e_1^2=e_2^2=2$, and
$e_1\cdot e_2=-1$). 
The ${\bf Z}_3$ twist acts with three {\em fixed} points. One of them is
located at the origin, $\xi_0=0$, while the other two are located 
(up to immaterial torus identifications) at $\xi_\pm=\pm \sqrt{v/\sqrt{3}} 
{\widetilde e}^2$, where ${\widetilde e}^a$, $a=1,2$, generate the $SU(3)$
weight lattice ${\widetilde \Lambda}\equiv\{m_a 
{\widetilde e}^a | m_a \in {\bf Z}\}$ (note that $({\widetilde e}^1)^2=
({\widetilde e^2})^2=2/3$, and
${\widetilde e}^1\cdot {\widetilde e}^2=1/3$). The space-group selection
rules imply that, for instance, three-point couplings of twisted fields
$T_A$, $T_B$ and $T_C$, where $A,B,C=0,\pm$ label the corresponding fixed
points, are non-vanishing if and only if $A=B=C$ or $A\not=B\not=C\not=A$
\cite{space}.
The couplings of twisted fields coming from the same fixed point are
unsuppressed. However, the couplings of twisted fields coming from 
different fixed points are {\em exponentially} suppressed. In particular,
they are suppressed by an exponential factor $\exp(-cvM_s^2)$ \cite{space}, 
where $c$ is
a numerical coefficient of order 1. Thus, in the large volume limit, that
is, when $v\gg M_s^{-2}$, the ``off-diagonal'' twisted sector couplings are
exponentially small. In the closed string theory language this can be
understood from the fact that such couplings come from world-sheet
instantons arising from strings wrapping the corresponding 2-cycles in the
orbifold. The action of such world-sheet instantons is proportional to
the volume $v$ of these two-cycles, hence the above exponential
factor. However, we can also understand this from the field theory
viewpoint - the
overlap of wave-functions for the states localized at different fixed
points is exponentially (more precisely, Gaussian-like) suppressed with the
distance between the fixed points (which is proportional to
$\sqrt{v}$). In fact, this way of thinking about this point makes it clear why 
in non-perturbative orientifolds some of the aspects of twisted open
strings such as twisted sector couplings should mimic the corresponding
statements in conformal field theory of twisted closed strings.

{}The above discussion might have interesting phenomenological implications
for the unification via Kaluza-Klein thresholds. Thus, as was pointed out
in \cite{zura}, to have a meaningful unification prediction we must ensure
that higher loops are subleading compared with the leading one-loop
threshold corrections to the gauge couplings. This is by no means automatic
in these scenarios. Thus, the number of heavy KK modes propagating in 
loops $N\sim vM_s^2\gg 1$. The actual loop expansion parameter then is {\em
not} given by the low energy gauge coupling $\alpha/4\pi$. Instead, the
loop expansion parameter is $\sim N\alpha/4\pi\sim \lambda_s$, where
$\lambda_s$ is the string expansion parameter (which in the conventions of
\cite{TASI} is related to the string coupling $g_s$ via
$\lambda_s=g_s/4\pi$). In fact, even though the low energy gauge coupling
$\alpha$ is relatively small, the loop expansion parameter $\lambda_s\sim
1$. Then without any additional cancellations higher loop corrections to
the gauge couplings would be as large as the one-loop threshold
contribution, hence lack of unification prediction.

{}There is, however, a way to make unification via KK thresholds
predictive. Thus, as was pointed out in \cite{zura}, if we consider ${\cal
N}=1$ supersymmetric theories, the heavy KK modes have ${\cal N}=2$
supersymmetry in the sense that their spectrum as well as their couplings
to each other are ${\cal N}=2$ supersymmetric. Then non-renormalization
properties of ${\cal N}=2$ gauge theories beyond one loop result in
additional suppression of higher loop corrections to the gauge couplings
due to (partial) ${\cal N}=2$ cancellations so that higher loop effects are
subleading compared with the leading one-loop threshold correction. This
renders unification via KK thresholds predictive.

{}A way to obtain ${\cal N}=1$ supersymmetric gauge theories with 
${\cal N}=2$ supersymmetry at heavy KK levels is via certain orbifold
compactifications. In particular, one can consider generalized
Voisin-Borcea orbifolds defined as\footnote{Note that in scenarios with 
unification in the TeV-range the volume of K3 must be large in the string units
in order for the observed four dimensional Planck scale to be much larger 
than the string scale, the latter being identified with the TeV-range 
unification scale.}
\begin{equation}\label{Voisin}
 {\cal M}=(T^2\otimes {\mbox {K3}})/{\bf Z}_M~,
\end{equation}          
where the generator $g$ of ${\bf Z}_M$ acts crystallographically on $T^2$ 
by $2\pi/M$ rotation $gz_1=\omega z_1$, and it is a symmetry of K3 acting
on the holomorphic 2-form $\Omega_2$ on K3 via $g\Omega_2=\omega^{-1}
\Omega_2$. Here $z_1$ is the complex coordinate parametrizing $T^2$, and
$\omega\equiv\exp(2\pi i/M)$. Note that $M$ can only take values
$M=2,3,4,6$ (or else the action of $g$ on $T^2$ would not be
crystallographic). The Calabi-Yau three-fold ${\cal M}$, which is an elliptic
fibration of $T^2$ over the base ${\cal B}\equiv {\mbox{K3}}/{\bf Z}_M$,
has $SU(3)$
holonomy, so, say, Type I compactified on ${\cal M}$ has ${\cal N}=1$
supersymmetry in four dimensions (provided that the corresponding tadpoles
can be canceled). 
Now consider D5-branes wrapping the fibre
$T^2$. The four dimensional massless modes of the gauge theory in the 
world-volume of the D5-branes have ${\cal N}=1$ supersymmetry. However, the
heavy KK modes on $T^2$ actually come in ${\cal N}=2$ supersymmetric
multiplets, and, as was shown in \cite{zura}, interactions involving only heavy
KK modes are also ${\cal N}=2$ supersymmetric. This is, therefore, a
setup where we could consider unification via KK thresholds.

{}As we have already mentioned, at least one of the chiral generations must
come from the twisted open string sectors\footnote{As was pointed out in
\cite{NP}, non-perturbative twisted open string sectors arise in ${\bf
Z}_M$ orbifolds with $M=3,4,6$, but not with $M=2$.} 
localized at fixed points of
$T^2/{\bf Z}_M$. Now, let us assume for a moment that all chiral
generations arise from such twisted open string sectors. Then, as we pointed
out above, couplings between different generations would be exponentially
suppressed in the limit of large volume $T^2$. That is, mixing between
different generations would be unacceptably small contradicting the
observed CKM matrix. Thus, at least naively, 
having large threshold corrections seems
incompatible with the desired mixing between different chiral generations
if all of them arise in ${\bf Z}_M$ twisted open string sectors. 

{}{\em A priori} we could assume that two of the chiral generations arise in
untwisted (with respect to the action of the $g$ twist\footnote{Note that
in the case of orbifold K3
these generations could still come from open string sectors twisted with
respect to another twist $\theta$ which is the generator of the K3 orbifold 
group ${\bf Z}_{M^\prime}$ (here ${\mbox{K3}}=T^4/{\bf Z}_{M^\prime}$).}) 
open string sectors while 
one generation comes from a twisted open string sector to possibly avoid
the above difficulty with small mixing. However, in certain cases it might
be desirable to have all three generations coming from twisted open string
sectors. For instance, in the TSSM to ensure proton stability in
\cite{zura} a {\em generation-blind} 
discrete ${\bf Z}_3\otimes {\bf Z}_3$ gauge symmetry was
proposed. This discrete symmetry can be identified with the corresponding 
orbifold discrete symmetry, which would imply that all three generations
would have to arise in ${\bf Z}_M$ twisted open string sectors
($M=3$ in this case - see the second reference in \cite{zura} for
details). The question we would
like to address next is whether we can have large threshold corrections
with all three generations coming from fixed points in the fibre $T^2$.

{}Here we would like to point out that there indeed seems to exist such a
setup. In particular, in the above discussion we have made an implicit
assumption that the untwisted NS-NS $B$-field in the directions of $T^2$ is
trivial. Let us, however, consider the case where we have a non-trivial
$B$-flux on the fibre $T^2$. In this case the Kaluza-Klein spectrum
is modified, and the behavior of the corresponding threshold
corrections with the volume of $T^2$ can be quite different from the case
without the $B$-field. In particular, as we will see in a moment, if there
is a non-trivial $B$-flux, we can have {\em large} threshold corrections
even if the volume of $T^2$ is small compared with
$\alpha^\prime=M_s^{-2}$.

{}For the reasons which will become clear in the following, we 
will mostly be interested in the ${\bf Z}_3$ and ${\bf Z}_6$
cases, where the fibre $T^2$ must have the appropriate 
${\bf Z}_3$ symmetry. However, for
illustrative purposes we will discuss the case of a square torus
$T^2=S^1\otimes S^1$ (with identical radii of the two circles) with a
non-zero $B$-field as this simple example captures all the key points
relevant for the subsequent discussions. Thus, the metric on $T^2$ and the
$B$-field are given by
\begin{equation}\label{torus}
 G_{ab}=\left(\begin{array}{cc}
               v & 0\\
               0 & v
              \end{array}\right)~,~~~
 B_{ab}=\left(\begin{array}{cc}
               0  & b\\
               -b & 0
              \end{array}\right)~,   
\end{equation}
where $v$ is the volume of $T^2$, and the $B$-field is defined up to unit
shifts $b\rightarrow b+1$. Without the $B$-field the metric felt by the
open string sector is simply $G_{ab}$. In particular, the KK modes coming
from D5-branes wrapping $T^2$ have mass squared $M^2_{\bf m}\sim G^{ab}m_a
m_b$ (here ${\bf m}\equiv(m_1,m_2))$, 
where $G^{ab}$ is the inverse of $G_{ab}$. However, if the $B$-field
is non-trivial, the metric felt by open strings is given by (we are working
in the units $2\pi\alpha^\prime=1$)\footnote{For a recent discussion of
D-branes wrapped on tori with non-zero $B$-flux using the boundary state
formalism, see \cite{harvey}.}
\begin{equation}
{\cal G}_{ab}=G_{ab}-B_{ac}G^{cd}B_{db}=
        \left(\begin{array}{cc}
               v+b^2/v  & 0\\
               0 &  v+b^2/v
              \end{array}\right)~,
\end{equation}
and, in particular, the KK modes have mass squared 
$M^2_{\bf m}\sim {\cal G}^{ab}m_a m_b$, where ${\cal G}^{ab}=
(v/(v^2+b^2))\delta_{ab}$ is the
inverse of ${\cal G}_{ab}$. Note that if the volume of $T^2$ is small (that
is, $v\ll\alpha^\prime$), in the case with the $B$-field (such that $b\sim
1$) we have {\em light} KK modes with masses of order $\sqrt{v}/b\ll
M_s$. This is to be contrasted with the case without the $B$-field where
the lightest massive KK mode has a mass of order $1/\sqrt{v}\gg M_s$. Thus,
in the case with the $B$-field we have a large number of massive KK modes
below the cut-off (that is, string) scale $M_s$ which contribute into the
threshold corrections to the gauge couplings. In fact, the leading one-loop  
threshold contribution is of order $N^\prime\sim b^2 \alpha^\prime/v\gg 1$, the
number of massive KK modes below the cut-off $M_s$. 

{}The above discussion can be straightforwardly generalized to the case of
the most general $T^2$. 
In fact, we are going to give the general expressions for the
one-loop threshold corrections to the gauge couplings in the small volume
limit in the presence of a non-trivial $B$-flux. First let us consider
D5-branes wrapping $T^2$ with the space transverse to the branes being K3.
The corresponding gauge theory has ${\cal N}=2$ supersymmetry at both
massless and massive KK levels. Let the gauge group be $\bigotimes_i G_i$
with the low energy gauge couplings $\alpha_i(\mu)$ (here $\mu$ is the energy
scale at which the gauge couplings are measured). The gauge couplings in
this case are renormalized only at one loop, and are given by:
\begin{equation}
 \alpha_i^{-1}(\mu)=\alpha^{-1}+{{\widetilde b}_i\over
 2\pi}\ln\left({M_s\over\mu}\right)+{\widetilde \Delta}_i~,
\end{equation} 
where $\alpha\equiv g_s/2\nu$ (we will define $\nu$ in a moment)
is the unified gauge coupling at the string
scale $M_s$, ${\widetilde b}_i$ are the corresponding ${\cal N}=2$ one-loop
$\beta$-function coefficients, and the {\em threshold} corrections
corresponding to the massive KK modes are given by (here we assume that
$\nu\gg 1$)
\begin{equation}
 {\widetilde \Delta}_i={{\widetilde b}_i\over 4\pi}\xi^2\nu-
 {{\widetilde b}_i\over 4\pi}\ln\left(\nu\right)+{\cal O}(1)~.
\end{equation} 
Here $\xi$ parametrizes the subtraction scheme dependence. In particular,
the infra-red (IR) 
cut-off in the loop integrals is taken as $\xi\mu$, while the
ultra-violet (UV) cut-off is taken as $\xi M_s$.

{}In the above expressions the quantity $\nu$ is defined as
\begin{equation}\label{nu}
 \nu\equiv(M_s/2\pi)^2\sqrt{\det({\cal G}_{ab})}=
 (M_s/2\pi)^2\det(G_{ab}+(2\pi\alpha^\prime) B_{ab})/\sqrt{\det({G}_{ab})}~.
\end{equation}
Note that in the small volume limit $v\ll \alpha^\prime$ we have 
$\nu\approx b^2(\alpha^\prime/v)$. In the large volume limit the $B$-field
does not play any important role, and we recover the
usual result $\nu\approx (M_s/2\pi)^2 v$ \cite{zura}.

{}Next, let us consider the case where D5-branes are wrapping the fibre
$T^2$ in the compactification on a generalized
Voisin-Borcea orbifold (\ref{Voisin}).
In this case we have ${\cal N}=1$ supersymmetry at the massless level,
while the heavy KK modes are still ${\cal N}=2$ supersymmetric. The gauge
couplings are now given by
\begin{equation}
 \alpha_i^{-1}(\mu)=\alpha^{-1}+{b_i\over
 2\pi}\ln\left({M_s\over\mu}\right)+\Delta_i~,
\end{equation}
where $b_i$ are the ${\cal N}=1$ one-loop $\beta$-function coefficients for
the massless modes. The one-loop threshold corrections in this case are
given by\footnote{Here we ignore heavy string oscillator thresholds which
are ${\cal O}(1)$ in ${\cal N}=1$ theories but still non-vanishing. Note
that in ${\cal N}=2$ theories such thresholds are absent - see \cite{zura}
for details and references.}
\begin{equation}
 \Delta_i={\widetilde \Delta}_i/M~,
\end{equation}
where $M$ is the order of the generalized Voisin-Borcea orbifold.

{}Thus, in the presence of the $B$-field we can have large threshold
corrections to the gauge couplings, so that unification via KK thresholds
can occur at low scales as in \cite{dienes,zura}. The key point here,
however, is that this is also the case for compactifications on {\em small}
(rather than only large) tori (or, more precisely, orbifolds thereof). That is,
the mixing between the chiral generations coming from (non-perturbative) 
twisted open string sectors localized at the corresponding orbifold fixed 
points need {\em no longer} be exponentially suppressed. 

{}The above discussion of D5-branes wrapping $T^2$ with non-zero $B$-flux
has one loose end which we would like to tie up next. In particular, so far
we have been assuming that the the $B$-field is taking generic
values. However, at special values of the $B$-field, namely, at certain
{\em rational} values of $b$, we can perform a T-duality transformation
which maps a small volume $T^2$ to a large volume ``dual'' torus - for a
recent comprehensive discussion and references, see \cite{SW}. The
corresponding KK threshold corrections in the case of a large volume $T^2$
are still expected to be large, and, as we will see in a moment, they are
identical in this T-dual description to those in the language of the
original small volume torus. However, at least naively, there might seem to
be a puzzle as for a large volume $T^2$ we expect twisted open
string sector states to have exponentially 
small off-diagonal couplings, which is {\em
not} the same as what we have concluded for a small $T^2$ compactification.    
As we will explain in a moment, there is no puzzle here as T-duality is a
good symmetry in the perturbative (from the orientifold viewpoint) open string
sectors but is broken in the non-perturbative twisted open string sectors
whenever we have the corresponding off-diagonal couplings.

{}The reason why the above point is especially important is that if we
consider Type I/orientifold compactifications, the $B$-field {\em cannot}
take arbitrary values but must be quantized. Indeed, let $\Omega$ be the
orientifold action interchanging the left- and right-movers on the closed
string world-sheet. The NS-NS $B$-field is antisymmetric under the action
of $\Omega$. Taking into account that it is defined only up to unit shifts,
it is then clear that $b$ can only take two values consistent with the
orientifold action: $b=0$ and $b=1/2$. In fact, the massless closed string 
states corresponding to the $B$-field moduli are projected out by the
orientifold action which implies that $b$ cannot take continuous
values. However, it can take the above quantized values, so from now on we
will focus on backgrounds with $b=1/2$. (For more details on orientifolds
with non-zero $B$-field, see, {\em e.g.}, \cite{bij}\footnote{In \cite{bij}
it was assumed that the closed strings that couple to D-branes wrapping,
say, a 2-torus with $b=1/2$ satisfy the ``no momentum flow'' condition in
the directions of $T^2$, that is, these states have left- and right-moving
momenta satisfying $P_{La}=-P_{Ra}$. This is only one of the two 
{\em a priori} available
choices. In particular, the second choice corresponds to
imposing the condition $P_{La}=-{{\cal R}_a}^b P_{Rb}$, where the matrix
${\cal R}$ is given by ${\cal R}\equiv E^T E^{-1}$ (see the next paragraph 
in the main text for
notations). In the former case the closed string states that couple to
D-branes satisfy $m_a-B_{ab}n^b=0$, which, in particular, implies that the
winding numbers $n^a$ must be {\em even}. This, as explained in \cite{bij},
leads to the rank reduction for the Chan-Paton gauge group. On the other
hand, if we impose the condition $P_{La}=-{{\cal R}_a}^b P_{Rb}$, then the
corresponding closed strings satisfy $m_a=0$ with arbitrary winding numbers
$n^a$, and the metric felt by open strings is ${\cal G}_{ab}$. 
In this case the rank of the Chan-Paton gauge group would not be
reduced. In the context of perturbative orientifolds, however, there is
an additional consistency condition one needs to impose, namely, that
the branes and orientifold planes couple to each other in a consistent 
fashion. This would require that the orientifold plane couples to the
closed string states whose mass squared $M_{\bf n}^2\sim {\cal G}_{ab}
n^a n^b$. On the other hand, in the loop channel for the Klein bottle
the orientifold action implies that the corresponding momentum states
have mass squared $M_{\bf m}^2\sim G^{ab} m_a m_b$. Note that since the
loop channel is related to the tree channel via the $t\rightarrow 1/t$
modular transformation, the consistency requires that the metric $G^{ab}$
be the inverse of ${\cal G}_{ab}$, which is only 
the case for the zero $B$-field.
Thus, the aforementioned second choice does not seem to be consistent in the
context of perturbative orientifolds
with non-zero $B$-field. However, in the framework of 
non-perturbative orientifolds, which we are interested in here,  
this need not be the case. In particular, we will assume that the $B$-field is 
quantized, and the metric felt by open strings is ${\cal G}_{ab}$.
Such a setup is conceivable in F-theory where we do not have orientifold
planes but the B-field must be appropriately quantized \cite{vafa,bersh}.}.)

{}For the sake of simplicity we will consider the case of a square torus
with the metric and the $B$-field given by (\ref{torus}). We will assume
that $b=1/k$, where $k\in {\bf N}-\{1\}$. In this case, if we start from a
torus with small volume, via a T-duality transformation we can map it
to a torus with large volume. To describe this T-duality transformation,
let us introduce the following matrix (here we are working in the units
$2\pi\alpha^\prime=1$, and we are closely following the discussion in
\cite{rocek,SW}):  
\begin{equation}\label{E}
 E_{ab}\equiv G_{ab}+B_{ab}~.
\end{equation}
The T-duality group in the case of $T^2$ is $SO(2,2,{\bf Z})$ whose
elements can be described in terms of $4\times 4$ matrices
\begin{equation}
 \left(\begin{array}{cc}
               \alpha & \beta\\
               \gamma &  \delta
              \end{array}\right)~,
\end{equation} 
where $\alpha,\beta,\gamma,\delta$ are $2\times 2$ matrices with integer
entries, and satisfy the following constraints:
\begin{eqnarray}
 &&\gamma^T\alpha+\alpha^T\gamma=0~,\nonumber\\
 &&\delta^T\beta+\beta^T\delta=0~,\nonumber\\
 &&\gamma^T\beta+\alpha^T\delta=I~.\nonumber
\end{eqnarray}
Here the superscript $T$ stands for transposition, and $I$ denotes 
the $2\times 2$ identity matrix. The above T-duality element acts on 
$E_{ab}$ as follows:
\begin{equation}
 E\rightarrow E^\prime=(\alpha E+\beta)(\gamma E+\delta)^{-1}~.
\end{equation}
Note that we can write $E=vI+b\Sigma$, where $\Sigma$ is the $2\times 2$
antisymmetric matrix with $\Sigma_{12}=1$. Next, consider the T-duality
transformation with (recall that $k=1/b$)
\begin{equation}
 \alpha=I~,~~~\beta=0~,~~~\gamma=k\Sigma~,~~~\delta=I~.
\end{equation}
We will denote this T-duality transformation by $P$.
The corresponding matrix $E^\prime$ is given by: 
\begin{equation}
 E^\prime={b^2\over v} I - b \Sigma~. 
\end{equation}
Note that the T-duality transformation $P$ amounts to $v\rightarrow b^2/v$,
$b\rightarrow -b$. In particular, a small volume torus with the $B$-field
is mapped to a large volume torus with the opposite $B$-field. 

{}So far we have not said anything about what happens to D5-branes under
the T-duality transformation $P$. As we will see in a moment, D5-branes
wrapping the original $T^2$ transform into D5-branes wrapping the dual torus
under the transformation $P$. This is consistent with the discussion after
(\ref{nu}), in particular, the threshold corrections computed on the
original small volume torus are identical to those on the dual large volume
torus, which can be seen by recalling that the T-duality transformation $P$
acts as $v\rightarrow (2\pi\alpha^\prime b)^2/v$ (here we have restored the
appropriate factors of $2\pi\alpha^\prime$).  
 
{}To explicitly see that the transformation $P$ maps D5-branes
to D5-branes (and not, say, D3-branes), let us analyze the structure of $P$
in terms of the familiar T-duality transformations $S$ and $T$. The
$S$-transformation 
corresponds to taking $\alpha=0$, $\beta=I$, $\gamma=I$ and $\delta=0$, and
amounts to mapping a 2-torus with metric $G$ and zero $B$-field to another
2-torus (with zero $B$-field) whose metric is given by the inverse of $G$. 
This is just the usual ``$R\rightarrow 1/R$'' type of T-duality
transformation. On the other hand, the $T$-transformation corresponds to 
taking $\alpha=I$, $\beta=\Sigma$, $\gamma=0$ and $\delta=I$, and
amounts to unit shifts of the $B$-field (but does not affect the metric on
$T^2$). It is not difficult to see that the transformation $P$ can be
written as
\begin{equation}
 P=S T^k S~.
\end{equation}
Thus, under the first $S$-transformation D5-branes are mapped to D3-branes,
the subsequent $T$-transformations do not affect the dimensionality
of branes, and the last $S$-transformations maps D3-branes back to D5-branes.

{}As we see, in the T-dual picture we have D5-branes wrapping a large
volume torus. We must then explain how come the twisted open string sector
states localized at the fixed points of the corresponding ${\bf Z}_M$
orbifold of $T^2$ have unsuppressed off-diagonal couplings in the original
picture while in the dual picture we expect them to be exponentially small.
To understand this point better, let us review some facts about twisted
open string states in non-perturbative orientifolds discussed in detail in
\cite{NP}. First,
non-perturbative twisted open string states do {\em not} arise in the ${\bf
Z}_2$ twisted sectors. There are twisted open string states arising in the
${\bf Z}_4$ twisted sectors, that is, $\Omega g$ and $\Omega g^{-1}$
twisted open string sectors, where $g$ is the generator of ${\bf
Z}_4$. However, only {\em one} of the two fixed points of $g$ acting on
$T^2$ gives rise to such states, namely, that at the origin. As pointed
out in \cite{NP}, the other fixed point does not give rise to
non-perturbative twisted open string states as there is a {\em twisted}
half-integer $B$-flux stuck inside of the corresponding ${\bf P}^1$ along
the lines of \cite{Asp}. In fact, this is consistent with the fact that in
the ${\bf Z}_4$ case T-duality on the fibre $T^2$ is a good symmetry of the
corresponding background, which is unbroken by the non-perturbative twisted
open string states \cite{NP}. Thus, in the ${\bf Z}_4$ case we have no
aforementioned puzzle with the off-diagonal couplings of twisted open
string states as there are {\em no} such couplings to begin with - all
twisted open string states come from the fixed point at the origin of
$T^2$. Thus, using T-duality in this case we can map the original
compactification with a small volume $T^2$ to the dual compactification with
a large volume 2-torus.

{}However, the situation is quite different in the ${\bf Z}_3$ and ${\bf Z}_6$
cases. We will focus on the ${\bf Z}_3$ case as the ${\bf Z}_6$ case is
similar. Thus, in the ${\bf Z}_3$ case we have twisted open string states
arising at {\em all} three fixed points of $T^2/{\bf Z}_3$, and they do
possess off-diagonal couplings. In the original small volume torus
compactification these are unsuppressed, whereas in the dual large volume
torus compactification they would be exponentially suppressed, which would
lead to a puzzle\footnote{Note that there is no such puzzle in the closed
string sector as the closed string states (both twisted and untwisted) 
do not possess Yukawa or
quartic scalar couplings - such couplings are absent in the parent
oriented closed string compactification with ${\cal N}=2$ supersymmetry. 
However, in the twisted open string sectors we do have the corresponding 
couplings so that one does need to address this issue.}. 
The resolution of this point is that, as was originally
pointed out in \cite{NP}, T-duality is {\em not} a good symmetry of this
background. In particular, it is broken by non-perturbative twisted open
string sectors (albeit the perturbative (from the orientifold viewpoint)
open string sectors do possess the corresponding T-duality symmetry). This
can be seen from, say, the corresponding explicit ${\bf Z}_6$ 
examples constructed in \cite{NP}. Thus, for instance, if we consider the
non-perturbative orientifold corresponding to Type I on
${\mbox{K3}}=T^4/{\bf Z}_6$, it contains both D9- and D5-branes (the former
wrap K3, while the latter are transverse to K3 and are located at the same
orientifold point at the origin of K3). The perturbative (from the
orientifold viewpoint) 99 and 55 open sting sectors have identical spectra
as they should - T-duality is a good symmetry in these sectors. However, the
twisted 99 and 55 open string spectra are {\em different}. In fact, the
correctness of these spectra follows from Type I-heterotic duality plus
anomaly cancellation requirements. In particular, had the twisted 99 and 55
open string spectra been the same, the six dimensional gravitational
anomalies would not have canceled. Thus, T-duality is indeed broken by
twisted open string sectors. The physical reason for this can be
intuitively understood by noticing that the aforementioned
T-duality transformation $S$ in the orientifold language maps $\Omega$ to
$\Omega R (-1)^{F_L}$, where $R$ is the reflection on the fibre $T^2$
($Rz_1=-z_1$), and $F_L$ is the left-handed space-time fermion number
operator. The two of the three $T^2/{\bf Z}_3$ fixed points $\xi_\pm$ (that
is, those {\em not} located at the origin) are {\em not} invariant under
the action of the reflection $R$, hence the lack of T-duality symmetry in
such backgrounds\footnote{A breakdown of T-duality due to non-perturbative
effects was also argued in a different context in \cite{Aspin}. It would be
interesting to understand if there is any relation between the results of
\cite{Aspin} and the earlier observations in \cite{NP}.}. The upshot of
this discussion is that we {\em cannot} use T-duality transformation $P$ to
map a small volume $T^2$ compactification to a large volume dual 2-torus
compactification, and this avoids the puzzle with the off-diagonal
couplings between the twisted open string states in these
backgrounds. Practically, this means that if we start from a
compactification with a small volume fibre $T^2$, we are stuck with this
description where the $B$-field plays a crucial role for unification via
Kaluza-Klein thresholds.

{}Before we finish our discussion, we would like to make the following remark.
First, we can ask if we could understand the KK thresholds
without appealing to string theory at all - after all, computation of KK
thresholds without the $B$-field can be carried out without any reference
to string theory except when discussing the UV cut-off, which is
chosen to be $\sim M_s$. In the case with non-zero $B$-field we can also give
a purely field theoretic description. In particular, in the regime of a 
small volume $T^2$ with the $B$-field we have a description in terms of
the corresponding gauge theory on a {\em non-commutative} 
torus\footnote{As we have argued above, in the ${\bf Z}_3$ and ${\bf Z}_6$
cases we would have no choice but to stick to the non-commutative
description if we chose to work within the field theory language. In the
${\bf Z}_4$ case, however, T-duality is a good symmetry of the
corresponding background, and we could rewrite the theory in the
commutative language. More precisely, the corresponding gauge theory is
still described in terms of the compactification on a non-commutative torus
(since the $B$-field is still non-zero in the T-dual picture), but it is
well approximated by the commutative description as the volume of the torus
is large (this is, in a sense, equivalent to the zero string slope limit
$\alpha^\prime\rightarrow 0$). At any rate, the 
${\bf Z}_3$ and ${\bf Z}_6$ cases might be
appealing for solving the proton stability as well as neutrino mass problems
in the TSSM (where we have three chiral generations), see the second
reference in \cite{zura} for details. On the other hand, these discrete 
symmetries might be relevant (see the second reference in \cite{TSSM4}) for
gauge coupling unification in the recently proposed TSSM4 model
\cite{TSSM4}, where we have {\em four} chiral generations, and the
electroweak Higgs is identified with a fourth generation slepton. 
Another potential use for such discrete symmetries might be within
the context of (discrete) flavor gauge symmetries proposed in \cite{flavor}
as a solution to the problem of flavor changing neutral currents (for other
related works, see, {\em e.g.}, \cite{flavor1}). In fact, it is likely that
such (discrete) flavor symmetries would require that the corresponding
chiral generations come from non-perturbative twisted open string sectors
as it is difficult to imagine how they would arise in the perturbative
orientifold framework.}  
(see \cite{SW} and references therein)\footnote{Some aspects of 
non-commutative extra large 
dimensions (that is, those transverse to the branes on which the Standard 
Model fields are localized) have been recently discussed in \cite{ardalan}.} 
with the UV cut-off at $\sim M_s$. This
is the motivation for the title of this paper.     

{}I would like to thank Martin Ro{\v c}ek and Tom Taylor for
discussions. This work was supported in part by the National Science
Foundation. I would like to thank the string theory group at Harvard
University for their kind hospitality while this work was completed. I
would also like to thank Albert and Ribena Yu for financial support.

\end{document}